# 30 GHz Zeno-based Graphene Electro-optic Modulator

Christopher T. Phare[1], Yoon-Ho Daniel Lee[1], Jaime Cardenas[1], and Michal Lipson[1,2,*]

[1]School of Electrical and Computer Engineering, Cornell University, Ithaca, NY 14850, USA

[2]Kavli Institute at Cornell for Nanoscale Science, Cornell University, Ithaca, NY 14850, USA

*Corresponding Author: ml292@cornell.edu

Graphene has generated exceptional interest as an optoelectronic material [1,2] because its high carrier mobility [3,4] and broadband absorption [5] promise to make extremely fast and broadband electro-optic devices possible [6,7,8]. Electro-optic graphene modulators reported to date, however, have been limited in bandwidth to a few GHz [9,10,11,12] because of the large capacitance required to achieve reasonable voltage swings. Here we demonstrate a graphene electro-optic modulator based on the classical Zeno effect [13] that shows drastically increased speed and efficiency. Our device operates with a 30 GHz bandwidth, over an order of magnitude faster than prior work, and a state-of-the-art modulation efficiency of 1.5 dB/V. We also show the first high-speed large-signal operation in a graphene modulator, paving the way for fast digital communications using this platform. The modulator uniquely uses silicon nitride waveguides, an otherwise completely passive material platform, with promising applications for ultra-low-loss broadband structures and nonlinear optics.

Integrated graphene modulators to date, by nature of their electroabsorptive structure, carry fundamental tradeoffs between speed and efficiency. In these structures, graphene forms at least one electrode of a large capacitor; a voltage applied to this capacitor causes carriers to accumulate on the graphene sheet and gates the interband absorption of the graphene through Pauli blocking [14]. This change in absorption modulates the intensity of light travelling through the waveguide. Operation speed can be increased by using a thicker gate oxide, but the lower capacitance makes for a lower carrier concentration change with voltage and reduced efficiency.

We overcome this tradeoff by exploiting the Zeno effect, in which an increase in loss in a coupled resonator increases the system transmission by changing the condition for resonator coupling (Figure 1a). We design a resonator to be critically coupled for low losses. When losses are increased, the resonator then becomes undercoupled, increasing transmission through the bus waveguide. This effect has been used to create sensitive all-optical switches [13,15]. Here, we use a silicon nitride ring resonator above a portion of which we integrate a graphene/graphene capacitor to modulate the round-trip ring loss (Figure 1b). At 0 V bias, both graphene sheets in the capacitor are lightly doped and thus opaque, so the ring has high loss and is undercoupled to the bus waveguide. Applying a voltage to the capacitor dopes the graphene sheets heavily, causing their absorption to decrease as the Fermi level crosses half the incident photon energy. The ring, now substantially lower-loss, couples to the bus waveguide, decreasing the system's transmission, as predicted theoretically [16]. Sensitivity to ring loss and on-state insertion loss can be designed by choosing the ring-waveguide coupling constant (Supplementary Information Figure 1). We note that this mechanism is not simply ring-enhanced absorption modulation, as the ring has little circulating power when the graphene is lossiest, and the modulator has the least transmission when the graphene is nominally transparent. Instead, attenuation occurs via destructive interference at the coupling region, and this interference is modified by the voltage-controlled ring loss. This destructive interference is significantly more sensitive to changes in loss than an electroabsorption modulator.

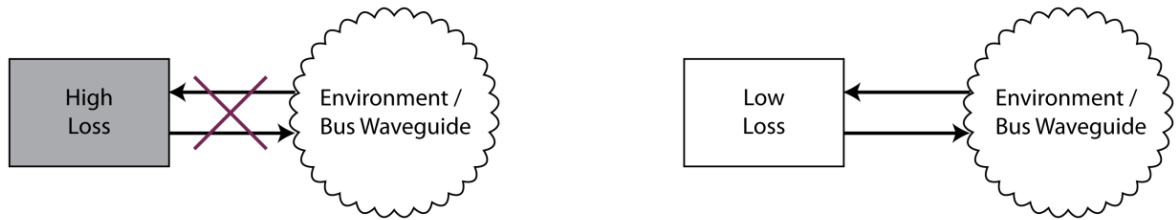

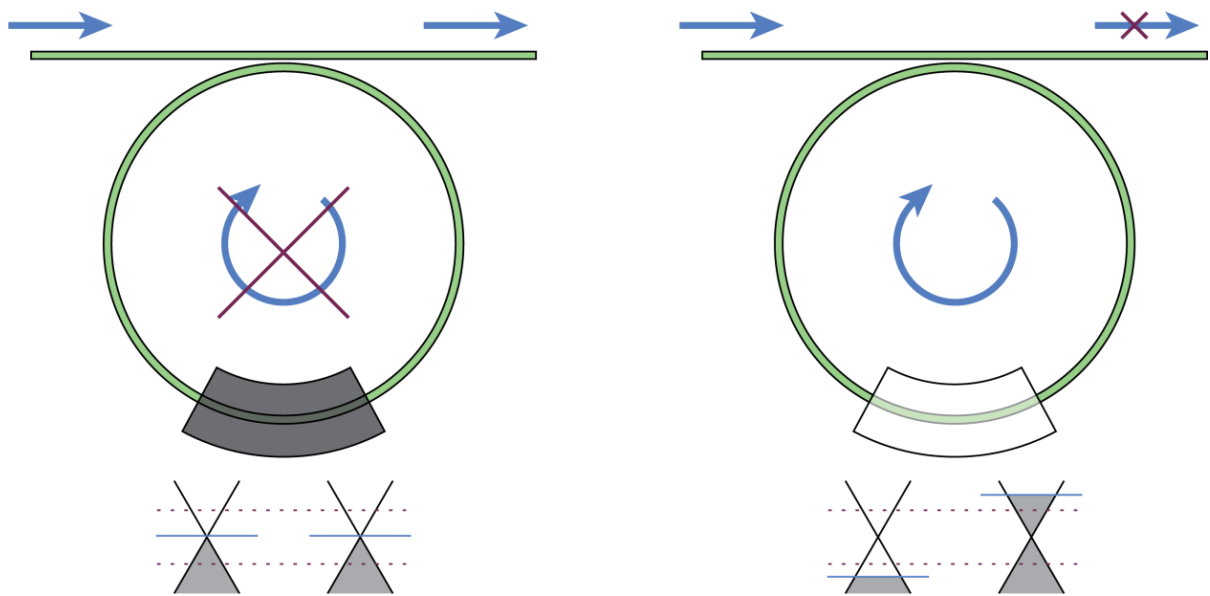

**Figure 1: Zeno effect. a.** Schematic concept. For a given coupling strength, a low-loss system will be more coupled to its environment than a high-loss system because of impedance matching. A resonator designed for critical coupling at low intrinsic losses would thus be undercoupled at high losses. **b.** Zeno effect in a graphene-clad ring resonator and band diagrams for the two gated sheets of graphene in the parallel-plate capacitor structure. When a graphene section with high loss is integrated with a ring resonator, it prevents light from circulating in the cavity, leading to high transmission through the bus waveguide. When electrostatically doped, the graphene becomes transparent, allowing light to circulate in the cavity and causing low transmission through the bus waveguide.

We demonstrate the Zeno effect by integrating graphene over a ring resonator fabricated from low-temperature plasma-enhanced chemical vapor deposition (PECVD) silicon nitride [17] (Figure 2).  In contrast to previously demonstrated graphene integration on silicon waveguides, which have relatively high loss and a limited transparency window, silicon nitride has ultra-low loss and broad transparency, enabling nonlinear and quantum systems.  Low-temperature deposition of silicon nitride also allows integration on the CMOS backend [18] or even on flexible substrates [19].  We use a waveguide cross section 1 μm wide by 300 nm tall to guide single-mode TE light.   This waveguide forms a ring resonator with radius 40 μm and bus waveguide coupling gap between 200 nm and 900 nm.  On top of a portion of the ring resonator we fabricate a graphene/graphene capacitor consisting of two sheets of monolayer graphene grown via chemical vapor deposition (CVD) on copper foil [20] and transferred via an optimized process to ensure cleanliness from metallic contamination [21] (see Methods).  Approximately 65 nm of atomic layer deposition (ALD) $Al_2O_3$ forms the interlayer dielectric.  The dielectric is five to ten times thicker than previous work [9,10], reducing capacitance and allowing our modulator to operate at much higher speeds.  The capacitor forms an arc along the silicon nitride ring with 30 μm optical path length and 1.5 μm width overlap between the two layers.  The graphene is completely encapsulated in $Al_2O_3$, providing an electrically insulating environment free from environmental effects or surface doping.  E-beam evaporated Ti/Pd/Au forms contacts to both layers of graphene [22].

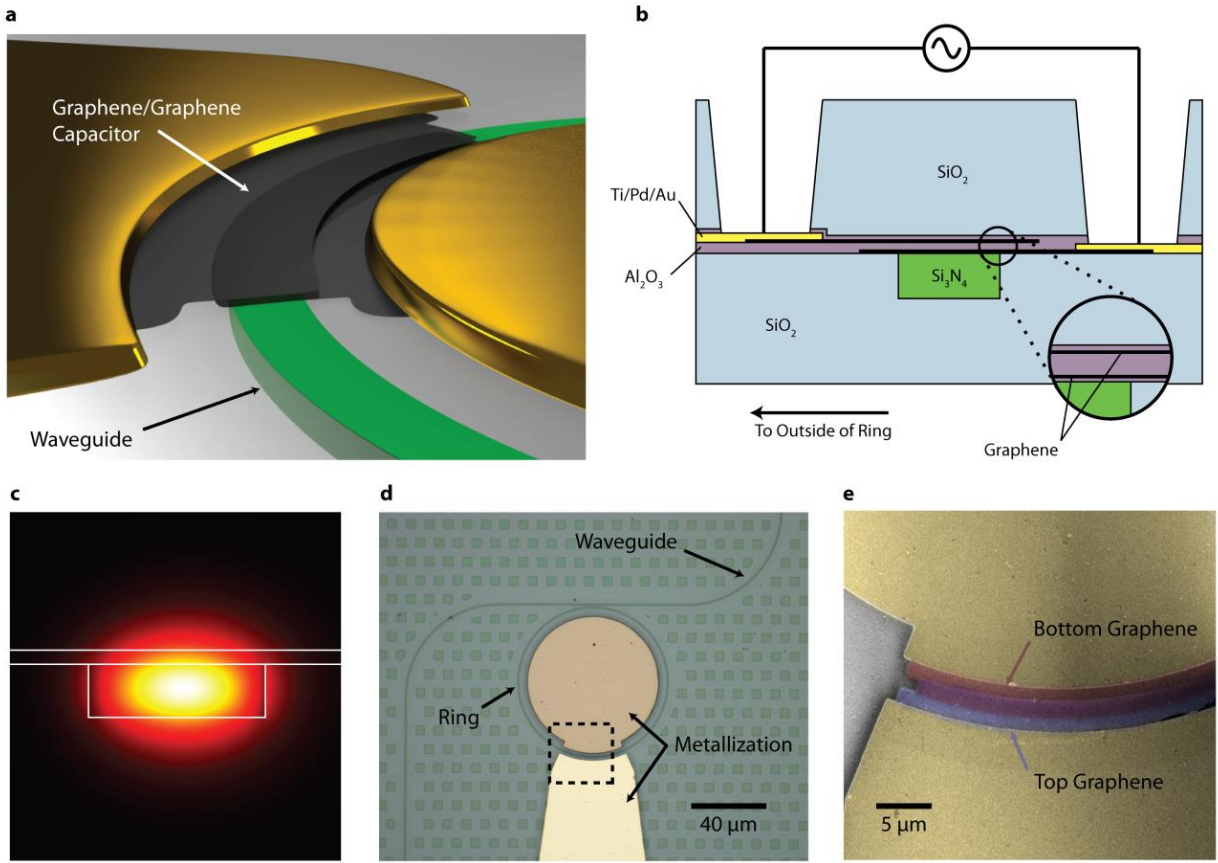

**Figure 2: Device Design a.** Schematic of the modulator consisting of a graphene/graphene capacitor integrated along a ring resonator. **b.** Cross-section of the device. Two layers of graphene separated by a 65 nm interlayer $Al_2O_3$ dielectric form a parallel-plate capacitor. **c.** TE mode Poynting vector showing boundaries of silicon nitride and $Al_2O_3$. The waveguide mode overlaps both graphene sheets. **d.** Optical micrograph showing bus waveguide, ring resonator, and Ti/Pd/Au metallization. Green squares are chemical mechanical planarization fill pattern. Scale bar, 40 μm. **e.** False-color SEM of dashed area in (d). Top and bottom graphene layers (blue and red) overlap in a 1.5 by 30 μm section over the buried ring waveguide. Gold areas indicate metal contacts. Scale bar, 5 μm.

We show the ability to tune the cavity transmission over 15 dB with 10 V swing (Figure 3a). Increasing the voltage (and thus decreasing the absorption in the cavity) changes the cavity lineshape from an undercoupled low-Q resonance at 0 V to a progressively more critically-coupled, higher-Q resonance. While the spectrum moves primarily vertically from changes in loss, shifts on the wavelength axis can be attributed to the non-monotonic gate-dependent imaginary conductivity of graphene and are similar to voltage-dependent cavity shifts for graphene integrated on photonic crystal cavities [23, 24]. Leakage current is below the measurement floor of the sourcemeter at all voltages, leading to negligible static power consumption.

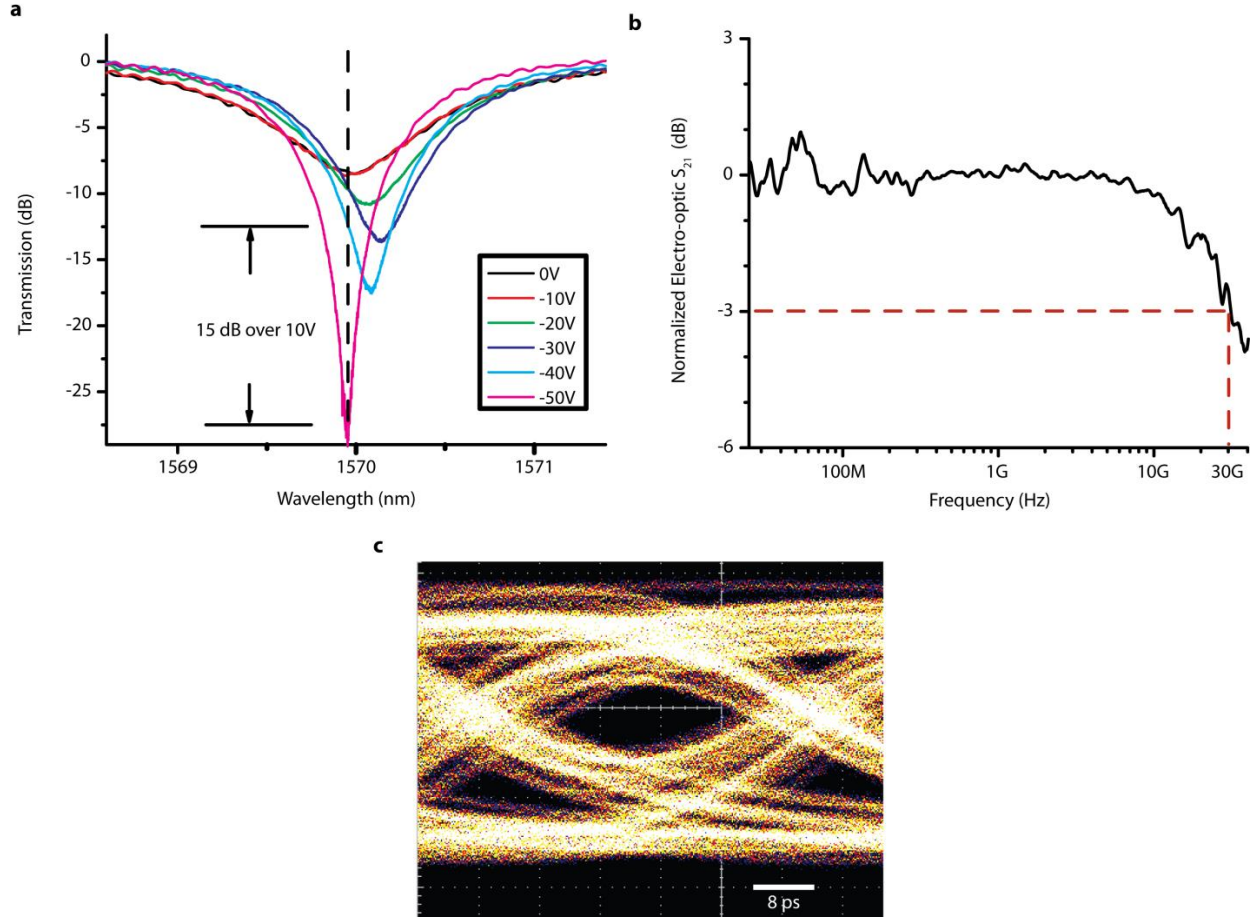

**Figure 3: Electrical Response a.** Transmission spectra for various applied DC voltages. The ring resonance sharpens and becomes critically-coupled for higher voltages, corresponding to lower losses in the graphene. **b.** Electro-optic $S_{21}$ frequency response. The device displays clear RC-limited behavior with 30 GHz bandwidth. **c.** Open 22 Gbps $2^7-1$ pseudo-random binary sequence non-return-to-zero eye diagram, measured at 7.5 V pk-pk and -30 V DC bias. Scale bar, 8 ps.

The device exhibits a small-signal RF bandwidth of 30 GHz, currently RC limited by the size of the capacitor, graphene sheet resistance, and graphene/metal contact resistance. We measure the transmission modulation with an electrical vector network analyzer and a 45 GHz photodiode by first tuning the laser to the center of the unbiased 1555 nm resonance, then increasing bias to -30 V. The photon lifetime in the ring resonator (130 GHz for Q=1500) does not limit the device bandwidth. For conservative estimates of the geometric capacitance (55 fF), roughly half of the (~100 Ω) resistance in the RC circuit comes from the 50 Ω transmission line itself. Thus, the intrinsic RC time constant, if driven, for example, by an on-chip source, is likely near 60 GHz. The remaining resistance is a combination of sheet resistance in the ungated graphene (~500 Ω/sq) and contact resistance (~500 Ω·μm), which we estimate via transfer length measurements (TLM) on the same growth of graphene. With current state-of-the-art contacts of ~100 Ω·μm [25, 26], the intrinsic speed of our device would approach 150 GHz.

To confirm the optical response of our modulator, we measure the large-signal response of the device and observe an open 22 Gbps non-return-to-zero eye diagram. The incoming signal is a $2^7-1$ PRBS at 7.5 V peak to peak without preemphasis and with a -30 V DC bias. To remove reflections caused by the strongly capacitive load of the modulator, we place, adjacent to the incoming signal probe, a second probe with a DC-block capacitor and 50 Ω RF termination. The bandwidth of our eye diagram is primarily limited by cabling losses and the 20 GHz bandwidth of the oscilloscope optical sampling module.

We have demonstrated the first ultrafast graphene modulator by leveraging Zeno coupling effects on a silicon nitride ring resonator. Such a dramatic improvement in the bandwidth of graphene modulators promises graphene's continued potential as an electro-optic material. Furthermore, the integration of a high-speed and broadband modulator with otherwise completely passive and broadly transparent waveguide materials opens many possibilities in nonlinear optics, quantum optics, and visible photonics.

**Methods**

We fabricated waveguides by depositing PECVD silicon nitride at 400 °C on 4 μm of thermally-grown silicon oxide and subsequently patterning with 248 nm deep-UV lithography. To provide a flat surface for the graphene transfer, which otherwise tends to break across the waveguide edge while drying, we deposit PECVD $SiO_2$ and planarize to the top surface of the waveguide using standard CMP techniques. CVD graphene on copper foil is spun-coat with 495 kDa polymethyl methacrylate (PMMA) in anisole, left to dry overnight, then floated on ferric chloride etchant for several hours and rinsed thoroughly in DI water. We then soak the graphene in dilute RCA-2 clean solution (hydrochloric acid, hydrogen peroxide, and DI water, 1:1:20), and rinse again prior to transfer. The wafer is coated with 10 nm thermal ALD $Al_2O_3$ and made hydrophilic by a short dip in stabilized piranha solution (Cyantek Nano-Strip) before the transfer. The graphene-coated wafer is left to dry, then baked at 145 °C for 15 minutes to relax wrinkles in the PMMA. We then remove the PMMA by soaking in acetone, rinsing with isopropanol, and baking at 170 °C for 10 minutes to ensure good adhesion with the substrate. The transferred graphene is patterned by deep-UV lithography and oxygen plasma. We thermally evaporate 1 nm of aluminum and allow it to oxidize in ambient air to serve as a seed layer for the ALD $Al_2O_3$ dielectric. We then lithographically pattern the metal layer and selectively remove the alumina with 30:1 buffered oxide etch immediately prior to e-beam evaporation and lift-off of 1.5 nm Ti / 45 nm Pd / 15 nm Au. The transfer, dielectric, and metallization processes are then repeated for the second layer. We then clad with a final $Al_2O_3$ layer and 2 μm of PECVD $SiO_2$, and open vias with reactive ion etching. Graphene remains somewhat p-type from the copper etch and wet transfer steps.

For small-signal RF measurements, we apply a DC bias of -30 V between bottom and top layers of graphene and an RF power of -17 dBm into a 50 Ω line. The coax line is connected to the device with a GGB model 40A microwave probe. We subtract losses from the cabling and bias tee, but not from the probe or photodetector (New Focus 1014), which has a nominally flat frequency response across the operating range of our device. The second termination probe is only used for eye diagram measurements. A detailed optical and electrical testing setup is shown in Supplementary Figure 2.

**Acknowledgements**

We thank Melina Blees and Paul L. McEuen for discussions and graphene samples. This work was supported in part by the National Science Foundation through CIAN ERC under Grant EEC-0812072. C.T.P. acknowledges support from a National Science Foundation Graduate Research

Fellowship under Grant DGE-1144153.  This work was performed in part at the Cornell NanoScale Facility, a member of the National Nanotechnology Infrastructure Network, which is supported by the National Science Foundation (Grant ECCS-0335765).  This work also made use of the Cornell Center for Materials Research Shared Facilities which are supported through the NSF MRSEC program (DMR-1120296).

**Author Contributions**

C.T.P. conceived the work, designed and fabricated the devices, and prepared the manuscript.  C.T.P. and Y.H.D.L. performed the high-speed measurements.  J.C. discussed device design and assisted with sample fabrication.  M.L. supervised the project and edited the manuscript.

# Supplementary Information

### I. Transmission vs. Ring Loss Characteristic Curves

Transmission on-resonance, neglecting phase effects, is described by

$$T = \left(\frac{a-t}{at-1}\right)^2$$

Where a is the round-trip transmission and t is the field coupling strength, controlled by the gap between ring and waveguide (t = 1 indicates no ring-waveguide coupling). Graphene's loss and any residual loss from the waveguide set (and modulate) a, while we can choose t arbitrarily to set a tradeoff between insertion loss (1-T) and voltage sensitivity ($dT/da \cdot da/dV$). Higher-Q rings, from lower residual loss, make the system more sensitive to changes in a.

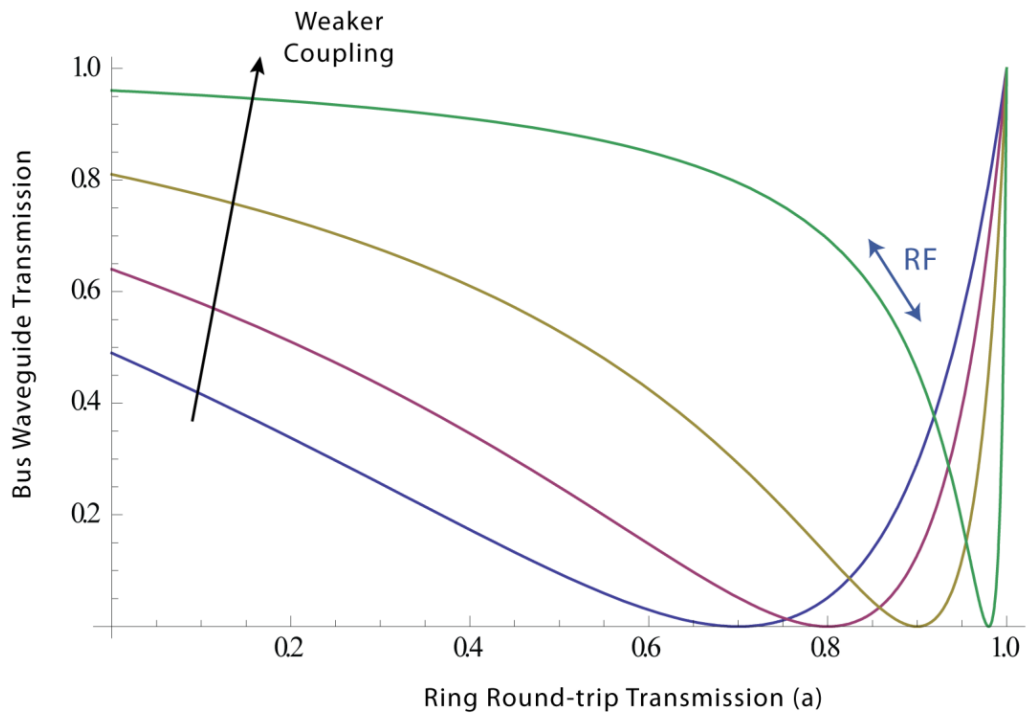

**Supplementary Figure 1: Efficiency Contours** The ring transmission on-resonance plotted for t = {0.7, 0.8, 0.9, 0.975}. Full extinction occurs at critical coupling (a = t), and modulation occurs on the negative slope for a < t. For a given achievable a (limited by graphene length and waveguide losses), choosing t makes a tradeoff between slope (efficiency) and transmission (insertion loss).

## II. Eye diagram test setup

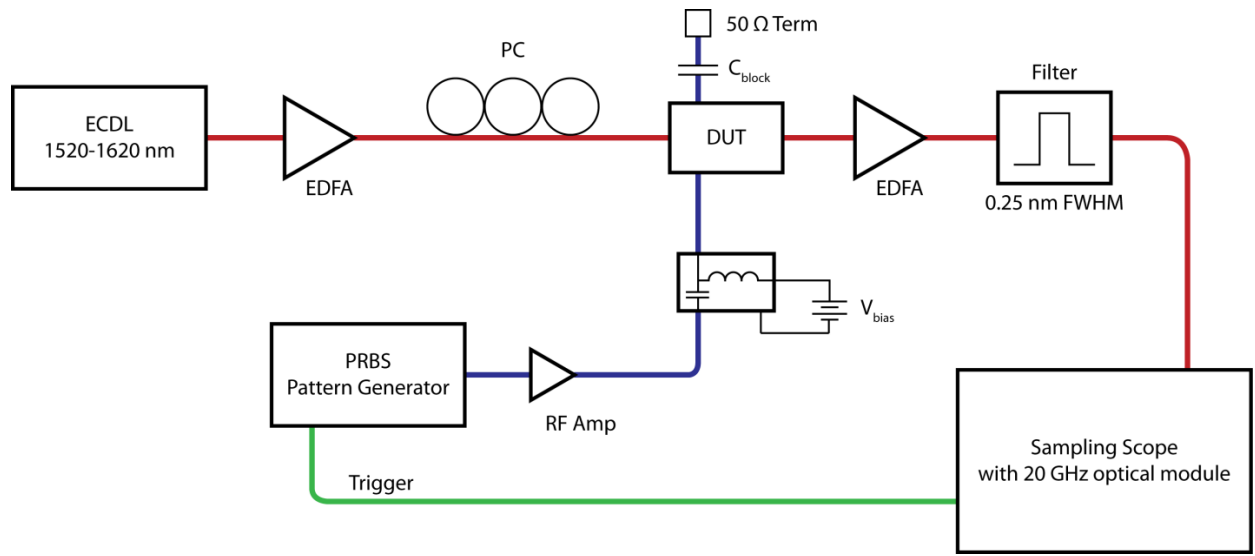

**Supplementary Figure 2: Eye Diagram Test Setup** The testing setup consists of an optical (red) and electrical (blue) arm. The external-cavity diode laser is amplified twice to compensate for chip facet and fiber component insertion losses, then sent through a bandpass grating filter to remove amplified spontaneous emission noise. RF PRBS signals (Centellax TG1P4A) are amplified in a 40 Gbps modulator driver (Centellax OA4MVM3) and biased before contacting the device with a GGB 40A picoprobe. A second probe blocks DC bias and terminates the RF signal to avoid reflections from the device.